\documentstyle[11pt]{article}
\def\lsim{\mathrel{\rlap{\raise 2.5pt \hbox{$<$}}\lower 2.5pt}}
\def\gsim{\mathrel{\rlap{\raise 2.5pt \hbox{$>$}}\lower 2.5pt}}
\begin{document}
\bibliographystyle{plain}
\thispagestyle{empty}
\begin{small}
\begin{flushright}
IISc-CTS-12/97\\
hep-ph/9712206\\
\end{flushright}
\end{small}
\vspace{-3mm}
\begin{center}
{\Large
{\bf
Lectures on Unification}}
\vskip 0.5cm
B. Ananthanarayan\footnote{Lectures given at the 
Sixth ICTP - BCSPIN Summer School
on Current Trends in High Energy Physics
and Cosmology, 
Kathmandu, Nepal, May 19 - June 3, 1997}\\
Centre for Theoretical Studies, \\
Indian Institute of Science, \\
Bangalore 560 012, India.\\

\vskip 3cm

\end{center}
\begin{abstract}
In these lectures we review the motivation, principles of
and (circumstantial) evidence for the program of unification
of the fundamental forces.   In an appendix, we review the
group theory pertinent to the program.
\end{abstract}

\newpage

\section{Introduction}

The standard model of the electromagnetic, weak and strong
interactions 
is the corner stone of elementary
particle physics\cite{gsw,cl}.  It is a lagrangian field theory
of quark, lepton and gauge bosons degrees of freedom
with the spontaneous breakdown of electroweak symmetry
achieved by an elementary higgs scalar potential.
While the standard model is enormously successful at
present day energies, it is likely to be the ``low''-energy
limit of a more complete and perhaps simpler description
of these interactions --- a description which derives
from the experience of the standard model, in the sense
of being a lagrangian field theory, being a gauge theory
and which uses the key concepts of spontaneous symmetry
breaking.   Grand unification\cite{azee,gg,ps,fm}, 
where in the standard
model gauge symmetry is embedded in larger symmetries
is such a program and is
a subject of the present set of lectures.  Another key
unification is that of particle types, viz., particles
of bosonic and fermionic types.  Supersymmetry\cite{mj,hpn,ms} is the
symmetry that treats these degrees of freedom on the
same footing and may be combined with lagrangian field
theory.  In particular, modern approaches to unification
simultaneously require grand unification as well as
unification of bosonic and fermionic statistics and is
called supersymmetric
unification and is the framework within which the present
discussion will take place.
These symmetries, however,
must be broken or hidden since there is no (direct)
evidence for such unification. 

In these lectures we will review the motivation, principles
and circumstantial evidence for the program of unification of
the fundamental forces, with the exception of the gravitational
forces.   
The aim of the lectures at this school is to bring
the participants up to date with the current status of research
in the areas covered at the school assuming as little as possible.
We will  mention virtually all the central notions
that enter the construction of the unification program, in 
{\it italics}.
We note, however that many of
the preliminaries are already presented in standard
textbooks\cite{ggr,rnm} and we will frequently refer the reader to them 
for tracing the primary sources.   
The relevant group theory is presented in an appendix
and is a summary of results discussed elsewhere\cite{rnc,rns}.
 
\section{The Standard Model}

At all length scales probed thus far at high energy accelerators,
there has been no evidence to suggest that the fundamental
constitutents of matter, namely the $quarks$ and the $leptons$
are anything but point-like.  The quarks come in the varieties
of up, down, charm, strange, top and bottom and the leptons
come in the varieties of electron and its neutrino, the muon
and its neutrino and the tau lepton and its neutrino.  
Of these the leptons do not participate in the {\it strong
interactions} and the neutrinos alone are {\it electrically} neutral.
Furthermore, the {\it weak interactions} are known to violate
$parity$, in that the left- and right- {\it chiral projections} of
these particles do not participate in the weak interactions
on par.   The quarks themselves are never seen isolated
in nature and are $confined$ to reside in hadronic matter
although at very high energies and on very short time scales there
is indubitable evidence for their existence.

All the forces listed so far result from the exchange
of $vector$ bosons, viz., quanta of fields that transform
as vectors under the Lorentz transformations.  
The vector particles themselves are introduced
via the {\it gauge principle}:  the gauge principle dictates
that the underlying Lagrangian field theory for the
interactions is invariant under {\it gauge transformations}
of the $local$ kind which in turn implies the existence
of a {\it covariant derivative},
schematically written as $\partial_\mu-ig A_\mu$,
which brings in the
vector fields of interest.  
The number of gauge fields
fields is equal to the number of {\it infinitesimal
generators} of the gauge symmetry.
The $photon, \ (\gamma)$ responsible for the long range 
electromagnetic interactions based on the
symmetry $U(1)$, the one-dimensional {\it unitary group}
is massless and exists in asymptotic states. 
On the other hand the
weak interactions which are short range are mediated by
the exchange of massive vector particles, the $W^{\pm}$ and
the $Z^0$.   
Finally the strong interactions mediated
by the massless $gluons, \ g$ rendered short ranged by a yet to
be discovered mechanism for the confinement of {\it colour}
quantum number that is carried by the gluon (indeed, as it
is by the quarks).  The gluons are the gauge bosons
of the underlying $SU(3)$ colour gauge group and eight
fields have to be introduced corresponding to the number
of infinitesimal generators.
The quarks come in three colors and transform as
triplets under the color gauge group, whereas the
leptons are singlets under this gauge group and
do not participate in the strong interactions at
the tree-level.  [In the following we will be
suppress the color indices assuming that they
are correctly summed over.]

The manner in which particles
are coupled to the gauge fields is dictated by which
{\it representation} of the relevant gauge groups 
they lie in.  The principle of gauge invariance
also dictates the manner in which particles interact
between themselves.  Only those couplings between the
particles are allowed which are left invariant by
the action of a gauge transformation. 
This picture, thus,  requires
us to specify the transformation properties of the
matter fields under the gauge group $SU(2)\times U(1)$.
In particular, the left-handed projections of the
u and d quarks, $q_L^T \equiv [u_L\ d_L]^T$ transforms as a doublet
under $SU(2)$ and carries the hypercharge $1/3$,
whereas the the right handed projections $u_R$ and
$d_R$ transform as singlets under $SU(2)$ and
carry hypercharges $4/3$ and $-2/3$ respectively.
Mathematically this would correspond to a term
in the Lagrangian density that would look like:
\begin{eqnarray}
& \displaystyle
\bar{q}_{L} i \gamma^\mu (\partial_\mu-\frac{ig}{2} T^i A_\mu^i
-\frac{ig'}{6} B_\mu) q_{L}+\bar{u}_R i(\partial_\mu-
\frac{2}{3} ig' B_\mu)u_R+\bar{d}_R i(\partial_\mu+
\frac{1}{3} i g' B_\mu) d_R &
\end{eqnarray}
Analogously we have the lepton doublets $[\nu_L \ e_L]^T$
which tranform as a doublet and with hypercharge $-1$
whereas the right-handed projections $\nu_R$ and $e^R$
transform as singlets and carry hypercharge of $0$ and
$-2$ respectively.  We note here that the right handed
neutrino is completely inert with respect to the
standard model gauge group and may even be left out
of the spectrum.  

A consistent picture arises when the
electromagnetic and weak interactions are considered
simultaneously in an electroweak framework based
on a group $SU(2)\times U(1)$ [where $SU(2)$ (or more
generally $SU(N)$ is the group of $2\times 2$ (or
more generally $N\times N$) unitary matrices] which
is then {\it broken spontaneously} by the {\it Higgs mechanism}
when a standard model Higgs doublet of scalar fields
$\phi^T=[\phi^+ \ \phi^0]^T$ is introduced
to produce $U(1)$ of electromagnetism, and in the process
turns three of the gauge bosons, now named $W^{\pm}$
and $Z^0$, massive.   The higgs mechanism occurs
when a quartic potential is introduced for the doublet
and when the classical potential turns into one
by the arrangement of specific relations between the
mass parameter and the quartic coupling wherein the
ground state is the asymmetric minimum.  More precisely
the higgs potential is written down as:
\begin{eqnarray}
& \displaystyle
{\cal L}_\phi=|(\partial_\mu -ig {T^i A^i_\mu \over 2} - 
{ig' \over 2} B_\mu) \phi|^2 -\mu^2 \phi^\dagger \phi -\lambda
(\phi^\dagger \phi)^2 & 
\end{eqnarray}
These transformation properties then specify the
nature of the kinetic energy terms of the standard
model particles, as we saw for the quarks, leptons
and the higgs fields.  
Finally the kinetic energy terms of
the gauge bosons themselves:  
\begin{eqnarray}
& \displaystyle -\frac{1}{4} F^i_{\mu\nu} F^{i \mu\nu}
-\frac{1}{4} B_{\mu\nu} B^{\mu\nu}
& \nonumber \\
& \displaystyle F^i_{\mu\nu}=\partial_\mu A^i_\nu -
\partial_\nu A^i_\mu + g \epsilon_{ijk} A^j_{\mu} A^k_{\nu},
B_{\mu\nu}=\partial_\mu B_\nu - \partial_\nu B_\mu & 
\end{eqnarray}
For the
non-abelian gauge fields the kinetic energy
involves self-couplings
of the gauge fields, a feature not present in
electrodynamics.  This has a crucial implication
for the strong interactions --- aymptotic freedom
is a property arising from this feature.
We must also note that the parameters in the Lagrangian
above themselves are ``running'' coupling constants whose
evolution is governed by the renormalization group equations.
In particular, for the gauge couplings, in the
standard model, we have the one-loop evolution
equations for the couplings:
\begin{eqnarray}
& \displaystyle
\frac{d\alpha_s^2}{d \ln Q^2} = \frac{1}{4 \pi} [11 - 4F/3] \alpha_s^2 & 
\nonumber \nonumber \\
& \displaystyle
\frac{d\alpha_g^2}{d \ln Q^2} = \frac{1}{4 \pi} [22/3 - 4F/3] \alpha_g^2 & 
\label{evolution} \\
& \displaystyle
\frac{d\alpha_{g'}^2}{d \ln Q^2} = \frac{1}{4 \pi} [- 20 F/9] \alpha_{g'}^2, & 
\nonumber
\end{eqnarray}
where $Q$ is the momentum and $F$ is the number of families.  We note here
that the the quadratic Casimirs of the representations
in which the gauge bosons and fermions enter the final
expressions since they represent the summing over the
intermediate particle states in the one-loop computation
of the beta functions.

Electroweak symmetry is broken when $\mu^2$ is chosen negative
with $\lambda>0.$  In particular, it is possible to arrange
the parameters to yield the vacuum expectation value to the
electrically neutral component: $\sqrt{2} v=[0 (-\mu^2/\lambda)^{1/2}]^T$. 
It is then possible to work through the details of the
higgs mechanism to produce expressions for the masses of
the $W^\pm$ and $Z$ bosons:  
\begin{equation}
m_W=g v/2, \, m_Z= \sqrt{g^2+g'^2} v/2
\end{equation}
One electrically neutral
scalar higgs boson is left behind after spontaneous symmetry
breaking.  Furthermore, from experiment, we have 
the relation for $v$ in terms of the Fermi constant $G_f=1.17
\cdot 10^{-5}$ GeV$^{-2}$, $v=2^{-1/4} G_F^{-1/2}=246.2$ GeV.

The transformation properties also
constrain the interactions between the particles
themselves; the term that we add to the Lagrangian
must be invariant under the combined gauge group.
Such terms imply Yukawa couplings between the higgs
doublet and left- and right-handed matter fields.
Gauge invariance allows terms of the type:
\begin{equation}
\bar{q}_L \phi q_R + {\rm H. C.}
\end{equation}
When spontaneous symmetry breaking occurs then
the vacuum expectation value of the Higgs field
multiplied by the Yukawa coupling gives rise to
an effective mass term for the standard model fermions.
Note that the absence of the right handed neutrino
implies that the neutrino is massless in the
standard model.  Furthermore, at the perturbative
level, the absence of lepton (and baryon) number
violating couplings rules out the possibility of
Majorana type masses which can be added to the
lagrangian to provide a mass to the left-handed
neutrino.
We also note that we would finally
have to sum over all the families.  Since all the
quarks have non-zero masses, once the electroweak
symmetry is broken, the quarks may mix amongst
themselves, viz., that the ``flavor'' eigenbasis
would now not correspond to the ``mass'' eigenbasis.
This would then be accounted for in the standard
model by the Cabibbo-Kobayashi-Maskawa mechanism
which is also rich enough to contain a single
CP violating phase.  We do not discuss this any
further except to note that the standard model
falls into the class of ``milliweak'' CP violating
models which is yet to be confirmed experimentally.

\section{Grand Unification}
A compelling goal of theoretical physics
is to replace what are the
engineering aspects of the standard model by a fundamental theory;
for example,
arbitrary parameters of the standard model, hitherto fixed
by experiment,  would then
be explained as consequences of a unified and symmetric
structure.  
Such a theory
would then make a whole host of predictions and simplifications
of our understanding of fundamental phenomena.  
Indeed, it would be very pleasing if the
seemingly arbitrary pattern of $SU(3)\times SU(2)\times
U(1)$ were to be aesthetically situated into an 
elegant framework.  
It is possible to envisage a scenario
wherein this is embedded in a larger group $G$, which would be
the basis of the gauge invariance of a theory manifest above
a unification scale $M_G$. 
The evolution of the standard model gauge couplings
does provide some credence to this belief as we
describe in one of the following subsections.
Below $M_G$, $G$ would be spontaneously
broken via the Higgs and possibility
some other mechanism to a sub-group large enough to
contain the standard model (in a multi-step scenario), which would
then be further broken down to the standard model gauge group
at various stages.  

\subsection{SU(5)}
Earliest examples of grand unification were provided by
those based on the groups $SU(4)\times SU(2)\times SU(2)$,
$SU(5)$ and $SO(10)$.  
In fact, the unitary group $SU(5)$
does admit the standard model gauge group as a {\it maximal
subgroup} and is an ideal candidate for unification.
Indeed, it is the smallest group large enough to
contain the standard model gauge group.  This
may be simply seen from erasing one of the
inner dots of the {\it Dynkin diagram} of
the Lie algebra of $SU(5)$. These
properties and other group theoretic
results maybe read off from Slansky's tables,
the essential mathematical steps recounted in the
book by Cahn and summarized in the appendix.

In this instance, we find that the standard
model gauge group's Lie algebra is a maximal subalgebra
of $SU(5)$ [$SU(4)\times U(1)$ being the other maximal
subalgebra], obtained by erasing one of the external
dots of the Dynkin diagram of $SU(5)$].  
Furthermore, when we consider the smallest
representations of $SU(5)$ namely the $\overline{5}$- and
10- dimensional representations.  Their branching
rules under the standard model gauge group $SU(3)\times
SU(2)\times U(1)$ are given by
\begin{eqnarray}
& \displaystyle (1,2)(-3)+(\overline{3},1)(2),\ {\rm and}
\, (1,1)(6)+(\overline{3},1)(-4)+(3,2)(1) &
\end{eqnarray}
respectively.  These may easily seen to be
precisely the quantum numbers of one 
standard model family.  In particular, they
corresond to the
left handed lepton doublet, right handed down-type 
quark singlet (conjugate), the right handed 
electron (conjugate),
the right handed up-type quark singlet (conjugate)
and the left handed quark doublet, respectively.
Among other things, this would imply that transitions
are possible between quark and lepton states [proton
decay problem] and mass relations between various
fermions, now unified into irreducible representations
of the groups.

It may also be seen that a $\overline{5}$- dimensional scalar
multiplet can accomodate the electroweak doublet
but the electroweak singlet, colored triplet must
be very massive in order to prevent rapid proton
decay\cite{hm}.  The 24 dimensional representation may also
be considered, with the branching rules:
$(1,1)(0)+(1,3)(0)+(3,2)(-5)+(\overline{3},2)(5)+(8,1)(0)$.
The singlet component is interesting, since a higgs
scalar in the 24- dimensional representation can be
used to break $SU(5)$ down to the standard model.
The 24- dimensional representation is also
the adjoint of $SU(5)$ which contains the gauge
bosons of the unified group.  Sure enough under
the standard model gauge group, we find candidates
for the electro-weak bosons, namely the $(1,1)(0)$
and the $(1,3)(0)$ and for the gluons, the
$(8,1)(0)$.  The rest must become supermassive
associated with the scale $M_G$.

The theory is specified by writing down the terms
in the Lagrangian that couple these fields.  In particular,
we see that Yukawa couplings may be written down for
the fermions in the $\overline{5}$ and $10$ and the
$\overline{5}$ dimensional scalar.  Indeed, one
may then compute the tensor products of these
irreducible representations and find in the sum
of irreducible representations a piece that is
invariant (i.e., a singlet) under $SU(5)$.  

\subsection{Charge Quantization}
The fact that standard model fermions of differing
hypercharges are accomodated into irreducible representations
of $SU(5)$ implies there is a basis for relating the
hypercharge assignments of those fermions that are
in the same multiplet.  For instance, when we consider
the electro-weak doublet and the down-type anti-quark
that lie in the same $\overline{5}$ it implies that
the action of the same diagonal hypercharge generator
produces eigenvalues of their respective hypercharges.
This in turn implies that charge is now quantized.
Furthermore, we have
the result that the normalization of
the hypercharge generator is now related to the normalization
of the diagonal generators of $SU(2)$ and $SU(3)$.
				     
The seemingly arbitary choice of gauge couplings
in the standard model would
also have to be replaced by a unique gauge coupling
in the event of unification into $SU(5)$.  
However, we must first fix the normalization
of the hypercharge generator of the standard
model, {\it vis a vis} the generator that is
embedded in $SU(5)$.  We recall the relations:
\begin{eqnarray}
& \displaystyle \sin^2 \theta_w=e^2/g^2=g'^2/(g'^2+g^2) &
\end{eqnarray}
In the standard model, we have the Gell-Mann-Nishijima
type relation:
\begin{eqnarray}
& \displaystyle Q=T_3+Y/2 &
\end{eqnarray}
However, in $SU(5)$, if we consider the
$SU(3)_c$ subgroup to lie in the upper $3\times 3$
diagonal sub-group and $SU(2)$ (weak-isospin)
to lie in the lower $2\times 2$ diagonal sub-group,
then $T_3={\rm diagonal}(0 \, 0\,  0\, 1\, -1)/2$ and
the hypercharge would be proportional to
$Y'={\rm diagonal}(-2\, -2\, -2\, 3\, 3)/(2 \sqrt{15})$.
If we have to correctly produce the electric
charge assignments to the $\overline{5}$, then
we would have to define
$Q=T_3+\sqrt{5/3}Y'$.  This then
gives us the required boundary condition
that $g'=\sqrt{3/5} g$ at the unification scale.

\subsection{Coupling Constant Unification}
A unification scale $M_G\sim 10^{16}
{\rm GeV}$ is suggested by gauge coupling unification,
above which physics would be described
by a grand unified theory\cite{azee}
based on a gauge group $G$.   
Indeed, the arrival at the structure
of fundamental interactions from renormalization group flow
has a predecessor in the example of asymptotic freedom in
deep inelastic scattering experiments and thus gauge
coupling unification is an extremely encouraging sign that
grand unified theories are the right step for a theory
of fundamental interactions.    
The evolution equations we consider are precisely
those we encountered earlier eq.(\ref{evolution}).
These equations provide the following system of
relations between the two inputs at low-energies
$\alpha$ and $\alpha_s$ and the unification scale,
the value of the unified coupling constant $\alpha_G$
and the value of $\sin^2\theta_w$ at low-energies.
\begin{eqnarray}
& \displaystyle M_G=Q_0 \exp (2\pi/11 \alpha) (1-8/3 (\alpha/\alpha_s)) & 
\nonumber \\
& \sin^2 \theta_w=1/6 + 5/9 (\alpha/\alpha_s) & \\
& 1/\alpha_G=3/8(1/\alpha -1/(6 \pi) (32/3 F -22) \ln M_G/Q_0. &
\end{eqnarray}
With the fairly accurately known inputs for
$\alpha=1/128$ and $\alpha_s=0.12$ at $Q_0=M_Z\sim 92 {\rm GeV}$,
we find the results $M_G\sim 1.2\cdot 10^{15} {\rm GeV}$,
$\sin^2\theta_w \sim 0.20$ and $\alpha_G\sim 1/41$.

\subsection{Complexity of Representations}
In the choice of gauge groups there are many theoretical
restrictions and furthermore in the choice of the representations
that could be of possible utility in model building.
One important property of the standard model that
singles out certain groups is the fact that the
weak interactions violate parity.  This implies the
existence of chiral fermions and the fact that left-
and right- handed chiral projections are assigned
to inequivalent representations of $SU(2)$.  When
viewed in the context of unification, this implies
that the representations we can use for accomodating
standard model fermions must be $complex$, where the
present notion of complexity implies that the image
of a group element in the representation and that of
its complex conjugate element cannot be made equal
by a similarity transformation using an element of
the representation.  It has been shown that the only
groups that admit complex representations are
$SU(n), n\geq 3, SO(4n+2)$ and $E_6$.

\subsection{$SO(10)$ and Anomaly Cancellation}
The seemingly arbitary assignments of a standard
model fermion to representations of $SU(5)$ finds
a natural resolution when we consider an even
larger gauge symmetry, viz., $SO(10)$.  It may be
easily seen from the Dynkin diagram structure of
the algebra of $SO(10)$ that $SU(5)$ is a subalgebra,
with $SU(5)\times U(1)$ being a maximal subalgebra.
We may either choose the $SU(5)$ as it stands as
the Georgi-Glashow $SU(5)$ or alternatively we
can choose a linear combination of one of the
diagonal generators of $SU(5)$ and the additional
$U(1)$ of the maximal subalgebra as hypercharge.
The latter corresponds to the so-called flipped
unification, wherein the word ``flipped'' refers
to the flipping of assignments of certain particles
to representations of the $SU(5)$, which we will
not discuss here.  $SO(10)$ is in a class of groups
that admit so-called spinor representations of
dimension of dimension 16 in this case.  The
branching rules for certain interesting and
important representations of $SO(10)$ under
$SU(5)\times U(1)$ read:
\begin{eqnarray}
& \displaystyle 10=5(2)+\overline{5}(-2) & \nonumber \\
& \displaystyle 16=1(-5)+\overline{5}(3)+10(-1) & \nonumber \\
& \displaystyle 45=1(0)+10(4)+\overline{10}(-4)+24(0) & \nonumber \\
& \displaystyle 126=1(-10)+\overline{5}(-2)+10(-6)+
\overline{10}(6)+45(2)+\overline{45}(-2) & \label{branchsu5}  
\end{eqnarray}
One may easily gather from here that the $16$- dimensional
representation in indeed the correct candidate for a standard
model generation and in addition contains a candidate for
a right-handed neutrino, which is an $SU(5)$ singlet.
The $10$- dimensional representation on the other hand
contains candidates for $SU(2)$ doublets that lie in the
$SU(5)$ 5- dimensional representations.
For pedagogical purposes we have also included the
branching rules of the $45$- of $SO(10)$ which would
contain the gauge bosons of $SU(5)$ and $U(1)$ which
might result for a direction in a scalar $45$- obtaining
a vacuum expectation value.  The branching rules
of the $126$- are given so as to provide a discussion
of Majorana masses for neutrinos in the following subsection.

Since the rank of $SO(10)$
[viz., the number of diagonal generators] is one larger
than the number of mutually commuting generators of
the standard model gauge group, it is possible to find
a $U(1)$ gauge boson, associated with the secondary
breakdown of the gauge symmetry $SU(5)\times U(1)$.
However, it is entirely likely that a single step
breaking of the gauge symmetry takes place in which
event it might be worthwhile to consider the branching
rules of the representation under the other maximal
subalgebra $SU(4)\times SU(2)\times SU(2)$:
\begin{eqnarray}
& \displaystyle 10=(1,2,2)+(6,1,1) & \nonumber \\
& \displaystyle 16=(4,2,1)+(\overline{4},1,2) & \nonumber \\
& \displaystyle 45=(1,3,1)+(1,1,3)+(15,1,1)+(6,2,2) & \nonumber \\
& \displaystyle 126=(6,1,1)+(\overline{10},3,1)+(10,1,3)+(15,2,2) &
\label{branchsu4} 
\end{eqnarray}
It would be instructive to think of the assignments of
the standard model fermions to the various multiplets of
$SU(4)\times SU(2) \times SU(2)$:  such a model is manifestly
left-right symmetric.  However, in order to be compatible
with phenomenology, it would be necessary to break one
of the $SU(2)$ and part of the $SU(4)$ down to $U(1)$ hypercharge
and $SU(3)$ respectively.  Here we also have the interesting
identification of the broken diagonal generator of $SU(4)$ with
lepton number.

Another outstanding feature of the standard model is the
possible appearance of gauge anomalies, associated with
triangle diagrams with axial vector currents at one of
the vertices of the triangle.  The assignments of hypercharges
in the standard model from phenomenology just serves
the purpose of cancelling the possible anomalies which
also calls in the presence of the color quantum number.
This mystery is not resolved even in the case of $SU(5)$
unification in which the particle assignments merely
rearrange the miraculous cancellation of the standard
model.  However the embedding of the gauge symmetry into
$SO(10)$ provides a {\it raison d'\^{e}tre} for
the cancellation.  This has to do with the fact that
in order to evalute the anomaly, one encounters the
following trace
\begin{eqnarray}
& \displaystyle {\rm Tr} \ \lambda^{ij}\{\lambda^{kl},
\lambda^{mn} \}
& \nonumber
\end{eqnarray}
where the $\lambda^{ij}(=-\lambda^{ji})$ represent the
generators of $SO(10)$ in a Cartesian basis.  This
result must necessarily be proportional to a 6-index
tensor which does not exist for any orthogonal group
with the exception of $SO(6)$.  Thus the representations
of $SO(n),n\neq 6$ are anomaly free.

\subsection{Neutrino Masses}
Note that whereas in the standard model, the field content
forbids a Dirac mass for the neutrinos since the right
handed neutrino is absent and Majorana mass is forbidden
by the conservation of lepton number.  In grand unified models,
neither of these principles is respected and a wide variety
of possibilities exists for the generation of neutrino masses.
However, far from being arbitrary, it should be possible to
uncover information regarding the structure of unified theories
from accurate determination of
small and eventually 
large neutrino masses and mixing angles,
{\it viz.}, neutrino masses may be viewed as bearing an imprint
on the structure of grand unification and the nature of the
breakdown of unification\cite{pm}.

One pedagogical example we consider is one wherein
the right-handed neutrino receives a Majorana mass of
the type $\nu^R\nu^R<126>$ when the $126$- dimensional
representation of $SO(10)$ receives a 
vacuum expectation value at a
supermassive scale, to break $SO(10)$ to
$SU(5)$.   This can be seen, when we consider
the branching rules under $SU(5)\times U(1)$, we find
that $126$ has an $SU(5)$ singlet component from
eq.(\ref{branchsu5}).  This gives the Majorana mass.
This Majorana mass is necessary to make the 
see-saw mechanism function to give a supermassive
mass to the right handed neutrino while making the
left handed component sufficiently light and preserving
mass relations for the Dirac masses.

It is commonly stated that the Majorana mass
must necessarily result from a $\Delta L=2$ vertex,
which means that the component acquiring the vev
must break lepton number.  This is seen by considering
the branching rules of the $16$ as well as the $126$
under $SU(4)\times SU(2)\times SU(2)$, wherein we
consider the first of the $SU(2)$ to be $SU(2)_L$.
The right handed neutrino lies in the $(\overline{4},1,2)$
while the direction of interest from the $126$ lies
in the $(10,1,3)$ component.  The branching rules
of the $10$ of $SU(4)$ under $SU(3)\times U(1)$
read $10=1(2)+3(2/3)+6(-2/3)$,  while that of the $\overline{4}$
reads $\overline{4}=1(-1)+3(1/3)$;  clearly one
may then have a Yukawa coupling $\overline{4}
\cdot \overline{4}\cdot 10$. The $SU(2)$ algebra
will admit a coupling between the 2- dimensional
representation in which the fermions are accomodated
and the 3- dimensional representation in which
the scalar lies.  Furthermore, when the $SU(3)$ singlet
direction in the 10- dimensional representation of
$SU(4)$ acquires a vev, lepton number is broken
proving the result that the Majorana mass requires
lepton number to be broken.

\subsection{Hierarchy Problem}
The presence
of disparate scales in the theory, $M_G$ and the weak scale $M_W
\sim 174\ {\rm GeV}$,
expected to be separated by more than ten orders of magnitude,
would render the mass of the Higgs scalar of the electro-weak model
$\sim M_W$, unnatural-natural.  Should the Higgs scalar be elementary, then
one manner in which it would remain naturally at the weak scale is due
to cancellation of divergences as in 
supersymmetric unified models\cite{mj,hpn}.
This is further discussed in the next section.

\section{Supersymmetric unification}

{\it This section is extracted from a recent
review article\cite{am} and is sufficiently detailed to serve
as a self-contained discussion of the subject.}

Supersymmetry is the unique symmetry that has non-trivial
commutation relations with the generators of the Lorentz group.
Supersymmetries enjoy non-trivial anti-commutation relations
amongst each other.  Their action on representations of the
supersymmetry algebra interchange the statistics between the
members.  Linear representations of the supersymmetry
algebra in relativistic field theory are realized in the Wess-Zumino
model\cite{ms}.   Important representations include chiral multiplets and
vector multiplets, which form the basis of the extension of the
standard model to various supersymmetric versions of the standard
model.  Since supersymmetry is not manifest in nature,
it must be broken, either spontaneously or explicitly.  It appears
that the second option is more favored, certainly more popular,
wherein supersymmetry is broken explicitly but softly.  The
requirement of soft supersymmetry breaking is in accordance with
the requirement of the well-known properties of supersymmetric
models including the cancellation of quadratic mass divergences for
scalars.  

In the context of grand unified model building, the existence of
scales $M_W$ and $M_G$ separated by several orders of magnitude
renders the mass of the elementary Higgs of the standard model
unstable and would drive it to the unification scale, without an
un-natural fine tuning of parameters of the Lagrangian.  The
cancellation of quadratic divergences in manifestly and softly-broken
supersymmetric theories renders supersymmetric versions of grand
unified models attractive candidates for unification.
The program of writing down a supersymmetric version of the
standard model, which is then embedded in a grand unified scheme,
[alternatively a supersymmetric version of a grand unified scheme]
may be realized by replacing every matter and Higgs field, by
a chiral superfield whose members carry the same gauge quantum numbers,
and by replacing every gauge field, by a vector super-multiplet.
Supersymmetry also requires that the standard model Higgs
doublet is replaced by two Higgs multiplets.  
This in turn leads to the introduction of another parameter
$\tan\beta$ which is defined as the ratio of the vacuum expectation
values of these two Higgs fields, $v_2/v_1$ where $v_2$ and $v_1$ 
are the vacuum expectation value of the Higgs fields that provide
the mass for the up-type quark and the down-type and charged leptons
respectively.
All the interactions
of the resulting model may then be written down once the superpotential
is specified.  Note that gauge invariance and supersymmetry allow
the existence of a large number of couplings in the effective theory
that would lead to proton decay at unacceptably large rates.
An ad hoc symmetry called R-parity is imposed on the resulting model
which eliminates these undesirable couplings and such a version has
received the greatest attention for supersymmetry search.
More recently models have been and are being considered where R-parity
is partially broken in order to study the implications to collider
searches.  
However such models are constrained by bounds on flavor
changing neutral currents as well as by the standard CKM picture,
also as it applies to CP violating phases.

In what follows we recall some of the 
essential successes of the recent investigations\cite{review} in
the theory of supersymmetric unification.
This was spurred by the confrontation of the ideas
of unification by the precision measurements of
the gauge couplings of the standard model at the
LEP\cite{adf}.  A highly simplified understanding of this
feature may be obtained from a glance at the one-loop
evolution equation for the standard model gauge couplings,
more correctly the gauge couplings of the minimal
supersymmetric standard model assuming that the effective
supersymmetry scale is that of the weak scale, with
$t=\ln \mu$:
$
\frac{d\alpha_i}{dt}=\frac{\alpha_i^2}{2 \pi} b_i, 
\ b_1=33/5,\ b_2=1,\ b_3=-3,
$
where we have assumed three generations.  One may then integrate
these equations to obtain:
$
\frac{1}{\alpha_i(M_Z)}=\frac{1}{\alpha_i(M_G)}+\frac{b_i}{2\pi}\ln
\frac{M_G}{M_Z}.
$
One may then use the accurately known value of $\alpha_{em}
(M_Z)=1/128$, with the identity $1/\alpha_{em}=5/3\alpha_1+1/\alpha_2$
which accounts for the normalization imposed by unification,
and the values of $\alpha_3(M_Z)\approx 0.12$ to solve for the
unification scale $M_G$ and the unified coupling constant $\alpha_G
\equiv 
\alpha_{1,2,3}(M_G)$.  One then has a prediction for $\sin^2\theta_w$
at the weak scale which comes out in the experimentally measured range.
Sophisticated analysis around this highly simplified picture up to
two and even three loops taking into account the Yukawa couplings
of the heaviest generation which contribute non-trivially at the
higher orders, threshold effects, etc., vindicate this picture of
gauge coupling unification which today provides one of the strongest
pieces of circumstantial evidence for grand unification\cite{deBoer}.

Predictions arising from (supersymmetric) unification
such as for the mass of the top-quark 
have been vindicated experimentally.
It turns out that unification based on $SO(10)$ 
is a scheme with great predictive power not merely in the
context of top-quark mass but also with implications
for the rest of the superparticle spectrum.  The primary requirement
that is imposed is that the heaviest generation receives its mass
from a unique coupling in the superpotential
$h{\bf 16.16.10}$
where the {\bf 16} contains a complete generation and
the complex {\bf 10} the two electroweak doublets\cite{als}.  When
the Yukawa couplings of the top and b-quarks and the $\tau$-lepton
are evolved down to the low energy and $\tan\beta$ pinned
down from the accurately known $\tau$-mass, one has a unique
prediction for the b and top-quark masses for a given value
of $h$.  If $h$ is chosen so as to yield $m_b(m_b)$ in its
experimental range, the top-quark mass is uniquely determined
up to these uncertainties.  
Now $\tan\beta\simeq m_t/m_b$ and the top-b hierarchy
is elegantly explained in terms of this ratio coming out
large naturally.
 
It is truly intriguing that
this picture yields a top-quark mass 
in its experimental range,
with $\alpha_S$ in
the range of the LEP measurements 
despite the complex interplay between the evolution equations
involved, the determination of the unification scale, running
of QCD couplings below the weak scale.  Note that this
requires that the top-Yukawa coupling must also come out
of order unity at $M_Z$.  It is also worth noting that
due to the nature of the evolution equations and competition
between the contributions to these from the gauge and Yukawa
couplings, this number $m_t(m_t)$ lies near a quasi-fixed point of its
evolution, {\it viz}, there is some insensitivity to the
initial choice of $h$\cite{sw}.  Moreover, if the $SO(10)$ unification
condition is relaxed to an $SU(5)$ one where only the b-quark
and $\tau$-lepton Yukawa couplings are required to unify at
$M_G$, $m_t(m_t)$ comes out in the experimental range
while preserving $m_b(m_b)$ in its experimental range for
$\tan\beta$ near unity.  In this event also the top-quark
Yukawa coupling lies near a quasi-fixed point which is numerically
larger compensating for the smaller value of $\sin\beta$
that enters the expression for its mass: $m_t=h_t \sin\beta 174
\ {\rm GeV}$.  Another interesting connection arises in
this context between the values of the Yukawa couplings at
unification and that of the gauge coupling when one-loop
finiteness and reduction of couplings is required:  such
a program also yields top-quark masses in the experimental
range\cite{kubo}.  

Besides the vindication of top-quark discovery predicted
by susy guts, another strong test takes shape in the form
of its prediction for the scalar spectrum.  In the MSSM 
the mass of the
lightest scalar is bounded at tree level by $M_Z$ since
all quartic couplings arise from the D-term in the scalar
potential.  The presence of the heavy top-quark enhances
the tree-level mass, but the upper bound in these models
is no larger than $140\ {\rm GeV}$.

Other predictions for softly-broken susy models arise
when a detailed analysis of the evolution equations
of all the parameters of the model are performed and
the ground state carefully analyzed.  In the predictive
scheme with $SO(10)$ unification, the model is further
specified by $M_{1/2},\ m_0$ and $A$, the common
gaugino, scalar and tri-linear soft parameters\cite{hpn}. 
It turns out that in this scheme $M_{1/2}$ is required
to come out to be fairly large, at least $\sim 500\ {\rm GeV}$
implying a lower bound on the gluino mass of a little
more than a TeV
and providing a natural explanation for the continuing absence of
observation of susy particles from scenarios based on radiative
electro-weak symmetry breaking\cite{als2}.
[An extensive study of the NMSSM with $SO(10)$ conditions
has also been performed\cite{ap}.]
Considerably greater freedom exists when the $SO(10)$
boundary condition is relaxed\cite{abs}.
In summary many predictions and consistency of the
MSSM and its embedding in a unified framework have been
vindicated;  however it is important to continue theoretical
investigations and checks to the consistency of these 
approaches and extensions
to include the lighter generations\cite{morereviews}.

\section{Gravitation}
The final frontier that still remains to be explored
is a framework within which a consistent incorporation 
of the gravitational
interactions is successful.  Whereas it has not been
possible to replace the Einstein theory 
by a quantum version due to bad ultra-violet
behaviour, supergravity possesses improved ultra-violet 
properties\cite{mj}.
String theories\cite{gsw2}
often contain supergravity in their low energy spectrum
and as a result supersymmetric unification is a favored candidate for
these reasons as well.

\bigskip

\noindent{\bf Acknowledgments:}  It is a pleasure
to thank the organizers and the Director,
Prof. Q. Shafi of the Sixth ICTP-BCSPIN
Summer School on Current Trends in High Energy Physics
and Cosmology for inviting me to present these
lectures..  I thank D. K. Ghosh and S. Kraml for careful reading
of the manuscript and helpful comments.

\newpage

\noindent{\bf {\Large Appendix: Review of Some Group Theory}}

\medskip

Much of the discussion presented below are those elements
of group theory required for unification.  Furthermore, our
discussion will only be confined to the theory of Lie algebras,
or in other words, the generators of the Lie groups of interest.
In what follows are the discussions of the Cartan subalgebra, the
roots of a Lie algebra, the Killing form, the metric on the algebra,
the notions of positive and simple roots, the Cartan matrix,
the translation to Dynkin diagrams, the restrictions on the
entries of the Cartan matrix, the notion of weights in the root-space
of a representation, the highest weight and the Weyl dimension formula.
Finally we present the formulae for computing the maximal sub-algebras
and the branching rules for representations, thus completing
the list of topics required for a discussion on unification.
The discussion closely follows that of Cahn and should be viewed
as a handbook to Slansky's Tables.

The language is developed for the algebra $SU(3)$ as an extension
of the familiar $SU(2)$ angular momentum algebra.  
$SU(2)$ is the group of $2\times 2$ unitary matrices and
is {\it homomorphic} to $SO(3)$, the rotation group in
three dimensions.  It is characterized by the three
operators $T_{1,2,3}$ and may be related to
the Pauli matrices and satisfy the commutation relations:
\begin{equation}
[T_i,T_j]=i \epsilon_{ijk} T_k
\end{equation}
Note that the group of $2\times 2$ unitary matrices
is obtained by 
\begin{equation}
\exp{i \theta_i T_i}.
\end{equation}
Furthermore, the existence of continuous symmetries
in field theory implies the existences of conserved
currents.

It is customary to define the combinations $T_{\pm}=
T_1\pm T_2$, which are the familiar raising
and lowering operators.  It is possible to diagonalize
only one of the operators $T_i$ and customarily it is
chosen to be $T_3$.  
In terms of these redefined operators, the
commutation relations now read:
\begin{equation}
[T_+,T_-]=2 T_Z, \, \, [T_3, T_\pm]=\pm T_\pm
\end{equation}
Furthermore, one defines the 
quadratic operator $T^2= (T^+\cdot T^-+T^-\cdot T^+)/2 + T_3^2$.
It is also well know that 
one may define a basis for higher angular
momentum states as an eigenstate of $T^2$ and
$T_3$ in terms of the quantum numbers $(j,m)$ with
$-j\leq m \leq j$ and the state is then $(2j+1)$-degenerate.
This is an example of a higher dimensional $representation$
of the angular momentum algebra.  
One may string out the entire $(2j+1)$-dimensional
basis vectors on a line and the lowering and raising
operators cause transitions between these states.

$SU(3)$ is
the smallest algebra which shows a structure rich enough to be
extended to the remainder of the {\it semi-simple algebras} namely the
{\it classical series} and the {\it exceptional series}.  
This algebra may be defined in terms of 8 linearly
independent operators.  
Conventionally these 8 may be represented by the
Gell-Mann matrices.
Of these, two may be simultaneously
diagonalized.  These simultaneously diagonalizable
which are called $T_z$ and $Y$ 
operators span the {\it Cartan sub-algebra} of the
original algebra.   These two are proportional to the
two diagonal Gell-Mann matrices $\lambda_3$ and
$\lambda_8$ respectively.  The remaining 6 operators
are named $T_\pm$, $V_\pm$ and $U_\pm$ and are equal to
$(\lambda_1\pm i \lambda_2)/2$, $(\lambda_4\pm i \lambda_5)/2$
and $(\lambda_6 \pm i \lambda_7)/2$ respectively.
We stick to these choices of linearly independent
operators since they naturally generalize the
raising and lowering operators of the $SU(2)$ algebra.

One may then work out the  commutation relations
between the 8 linearly independent operators knowing
their representations in terms of the Gell-Mann matrices.
For instance, we may list the commutation relations
enjoyed by $T^+$ with system of operators we have chosen:
\begin{eqnarray} \label{commrel}
& \displaystyle [t_+, t_+]=0, [t_+, t_-]=-2 t_z, [t_+,t_z]=t_+,
[t_+,u_+]=-v_+ & \\ \nonumber
& \displaystyle [t_+,u_-]=0, [t_+,v_+]=0, [t_+,v_-]=u_-,
[t_+,y]=0 &
\end{eqnarray}
Since we are working with a Lie algebra, if we take
an arbitary linear combination of our 8 operators and
consider its commutation relation with a fixed operator
out of the 8, we produce a different linear combination
of the original 8 operators.   Corresponding to each
of the 8 original operators $X_i$, we would find 8 different
linear combinations.  Our knowledge of linear algebra
teaches us that we may therefore represent each of these
by $8\times 8$ matrices, we call ${\rm ad}X_i$ and is called the   
{\it adjoint representation}.  By fixing an order for the operators $X_i$, 
one may produce explicit representations for ${\rm ad} X_i.$
In particular, if we fix the order of the $X_i$ to be
$T_+, T_-, T_z, U_+, U_-, V_+, V_-, Y$, the representation
for ${\rm ad}(a T_z+b Y)$ is an $8\times 8$ diagonal
matrix with the diagonal entries $a, -a, 0, (-a/2+b),
(a/2-b), (a/2+b), (-a/2-b), 0$.  The original choice
of the linearly independent basis is now justified;
each of them that is not in the
Cartan subalgebra is now called a {\it root vector} and
the corresponding diagonal entries is called the $root$.
Generalization to other algebras may be performed by
considering the operators that lie in the Cartan-subalgebra
and the remainder broken up into (dim G - dim H) root
vectors.  

It is now possible to associate to the adjoint representation
the {\it Killing form}:
\begin{equation}
(X_i, X_j)={\rm Tr}\ {\rm ad}\ X_i {\rm X_j}.
\end{equation}
It turns out that for our choice of linearly independent
vectors, the only non-zero answers occur for
\begin{eqnarray}
& \displaystyle (t_z,t_z)=3, (y,y)=4 & \\ \nonumber
& \displaystyle (t_+,t_-)=6, (v_+,v_-)=6, (u_+,u_-)=6 &
\end{eqnarray}

In short, we have the
root vectors $\alpha_i(k), i=1,2,3$ of the algebra with roots
$\pm a, \pm (-a/2+b), \pm (a/2+b),$ respectively when we chose
the vector in the Cartan subalgebra $k=a t_z + b Y.$
[Note that $\alpha_3=\alpha_1+\alpha_2$.]
Corresponding to these three roots are the vectors in
the Cartan subalgebra $h_{\alpha_i},\, i=1,2,3$,\,
$h_{\alpha_1}=t_z/3$, $h_{\alpha_2}=-t_z/6+y/4,$ $h_{\alpha_3}=t_z/6+y/4$
such that $\alpha_i(k)=(h_{\alpha_i},k)$.
Now, we may define the scalar product on the space
of roots with the definition:
\begin{equation}
<\alpha,\beta >=(h_\alpha,h_\beta).
\end{equation}
In particular, for the system of roots $\alpha_i$, we have
$<\alpha_i,\alpha_1>=1/3, <\alpha_1,\alpha_2>=-1/6,
<\alpha_1,\alpha_3>=1/6, <\alpha_2, \alpha_3>=1/6.$
This can be expressed geometrically as vectors of equal
length $1/\sqrt{3}$, with $\alpha_1$ and $\alpha_2$ at
an angle of $120^0$ and $\alpha_1$ and $\alpha_3$ at
an angle of $60^0.$
These may be represented as non-orthogonal vectors
in a two-dimensional plane.  Generalization to
higher algebras would entail the representation of
roots in a space whose dimension is equal to
dim H.

What we can observe from the eq.(\ref{commrel}) and
the definition of the Killing form and the structure of
the roots and the associated root vectors, is that
the commutation relation between root-vector $e_\alpha$ of
the root $\alpha$ and $e_{-\alpha}$ of the root $-\alpha$
yields the element of the Cartan algebra $h_\alpha$
multiplied by the Killing form $(e_\alpha,e_{-\alpha})$,
the commutation relation between a root-vector and an
element of the Cartan algebra produces an eigenvalue equation
for the same root-vector, where the eigenvalue is the root
in question and finally, a commutation relation between
two root-vectors yields an expression that is non-zero
only if the sum of the two roots associated with the root-vectors
is itself a root:  $[e_\alpha,e_\beta]=N_{\alpha\beta} e_{\alpha+
\beta}$, if $  \alpha+\beta$ is itself a root, or zero otherwise.

These properties may be simply generalized
for a larger and more abstract (semi-simple) Lie algebra.
However the generalization itself imposes severe
restrictions on the nature of the root-space.
In order to discuss the generalization, we will
first of all discuss the higher dimensional representations
of the $SU(3)$ algebra, having encountered thus far,
the fundamental representation and the adjoint
representation.    A representation is obtained
when we have for each element of the algebra
a linear transformation (i.e., a matrix)
on a vector space (i.e., column vectors) 
that preserves the commutation relations.
Note that for members of the Cartan subalgebra
we can simultaneously diagonalize the associated
matrices and the column vectors $\phi^a$ can be
so chosen such that
\begin{equation}
H_i \phi^a=\lambda_i^a \phi^a
\end{equation}
In the case of the fundamental representation of
$SU(3)$ with $T_z={\rm diag}(1/2, -1/2, 0)$ and
$Y={\rm diag}(1/3,1/3,-2/3)$, the weight vectors
are $\phi^a=[1,0,0]^T, \phi^b=[0,1,0]^T$ and
$\phi^c=[0,0,1]^T$, and with $H=a T_z+b Y$
we find 
$H \phi^a=(a/2+b/3) \phi^a=(2\alpha_1/3+\alpha_2/3)\phi^a$,
$H \phi^b=(-a/2+b/3) \phi^b=(-\alpha_1/3+\alpha_2/3)\phi^b$ and
$H \phi^c=(-2b/3) \phi^c=(2\alpha_1/3+\alpha_2/3)\phi^c$.
The eigenvalues of the eigenvalue equations $M^i, i=a,b,c$ above
are known as the weights of the representation and
the corresponding eigenvectors $\phi^i, i=a,b,c$
and are known as the
weight vectors corresponding to that weight.
We have also shown that it is possible to express
the weights as linear combinations of the roots.
It may then be shown that that image of the root
vectors $e_\alpha$ in the matrix representation $E_\alpha$
when acting on $\phi^a$, produces a weight vector
corresponding to the weight $M^a+\alpha$, unless
$E_\alpha \phi^a=0$.  Thus the $E_\alpha$ play the
role of raising operators and the $E_{-\alpha}$ as
lowering operators.  Just as we may represent
the root-vectors in a two-dimensional plane,
the weight-vectors may also be represented by
points on the same two-dimensional plane.
Another way of expressing this is to say that
the weight-vectors, in general, are a linear
combination of the root-vectors.
The action of the raising and lowering operators
associated with a fixed root would be take one
to another unless the action terminates, just
as in the case of the $SU(2)$ algebra the
action of the lowering operator would be
to cause transitions between the states
in a multiplet unless $m=-j$.   
In particular, if we have a weight $M$ that
lies in the string $M+p\alpha, ... , M, M-m\alpha$,
then the following relations hold:
\begin{eqnarray}
& \displaystyle m+p={2 <M+p\alpha,\alpha>
\over <\alpha,\alpha>} & \nonumber \\
& \displaystyle m-p={2 <M,\alpha>
\over <\alpha,\alpha>} & 
\end{eqnarray}
Just as
we have reflection symmetry about the orgin
in $SU(2)$ algebra, for larger algebras
we have a richer symmetry structure which is
known as the $Weyl\ Group.$ 

One may then use the defining
properties of Lie algebras to deduce many of the
properties of weight vectors in general.  
The multiplicity of states with a fixed
weight may in principle exceed one.  However
for the adjoint representation it is unity
for each root with the exception of the
Cartan sub-algebra.  Furthermore it turns
out that for the root-system, the following
identity has to be respected:
\begin{equation}
\frac{<\alpha,\beta>^2}{<\alpha,\alpha><\beta,\beta>}=\frac{mn}{4}
\end{equation}
where $m$ and $n$ are integers [this follows
from an important and interesting property
of the roots that if $\alpha$ is a root,
$2\alpha$ cannot be a root].  However, the left hand side may be
seen to be nothing but $\cos^2\theta$ where $\theta$
is the angle between the root-vectors $\alpha$ and $\beta$.
This then implies that $\cos^2\theta$ can be $0,1/4,1/2$ and $3/4$.

Now we describe further characteristics of the $SU(3)$ root
system which by now may have already become evident:
while there are 6 roots, 3 or them are negatives of
the other 3.  Of these only two are linearly independent.
One then considers a certain ordering of these roots
to define the notion of a $positive$ root.  In the
present case, these may seen to be $\alpha_1, -alpha_2
$ and $\alpha_3$ [simply put, these roots are the ones
where the coefficient of $a$ is positive when we consider
the commutation relations of the root vectors with
$aT_3+b Y$].  Out of these, $\alpha_1$ can be written
as $\alpha_3 + (-\alpha_2)$.  Then we are led to the
definition of a $simple$ root as one that cannot be
written as a sum of two positive roots.  In the
present example, $-\alpha_2$ and $\alpha_3$ are the
simple roots.

All the information regarding the algebra
can then be expressed in terms of the $Cartan$ $matrix.$
The Cartan matrix is defined as the matrix whose
elements are given by
\begin{equation}
A_{ij}={2<\alpha_i,\alpha_j>\over <\alpha_j,\alpha_j>}
\end{equation}
In the case of $SU(3)$, we see quite simply
that the diagonal elements are $2$ and the off-diagonal
elements are equal and each is $-1$.  $SU(3)$ belongs
to what is known as the classical series of algebras
and in particular to the one wherein all the simple
roots are of equal length [this property is called {\it simply laced}
and the off-diagonal elements are equal].  Detailed study of
the properties of the root-systems of algebras in
general also shows that simple roots can come in
atmost two lengths.  Thus the angles between the
roots and their lengths completely characterize
the algebra.  Given these the Cartan Matrix may
be written down for any algebra and must be
subject to the constraints of the root-system.
Besides the classical series on which there is
no restriction on the number of simple roots, viz.,
no restriction on the dimension of the Cartan subalgebra,
there is the $exceptional \ series$ all of which
are known.  It turns out that the root-systems of
the classical series are in one-to-one correspodence
with the algebra of the infinitesimal generators of
the unitary, orthogonal [of even and odd
order] and symplectic groups.  These are the
$A_n$, $B_n$, $D_n$ and $C_n$ series.  The
exceptional series consists of $G_2, F_4$,
$E_{6,7,8}$ which have been documented
in several references.  Of particular interest
to us will be the unitary, orthogonal and the
$E$-exceptional series.  

The Cartan matrix language for treating
the Lie algebras may be translated into what
are known as the Dynkin diagrams.
The Dynkin diagrams code the information
by representing the simple roots by dots
of two types [if the roots are of unequal
length] and the $i$ and $j$ roots are joined
by the larger number of the entries $A_{ij}$ or
$A_{ji}$.

The Dynkin diagram technique makes
it very simple to study the subalgebras
by erasing dots out of the Dynkin diagrams
(or their extended versions).   The extension
of the Dynkin diagrams in order to evaluate
the maximal subalgebras is performed using
standard techniques.

Corresponding to each representation, one may
define the $Dynkin$ label of the representation $\Lambda$:
\begin{eqnarray}
& \displaystyle \Lambda_i=\frac{2<\Lambda,\alpha_i>}{<\alpha_i,
\alpha_i>} &
\end{eqnarray}
Multiplication of the vector $\Lambda_i$
by the inverse of the Cartan matrix,
which is known as the {\it metric tensor} on the root
space, express the element of the representation as
a linear combination of the simple roots.

Given
the Dynkin label of the highest weight, one may
then evaluate the dimensionality of the representation
to which it belongs by use of the Weyl dimension
formula which reads:
\begin{eqnarray}
& \displaystyle {\rm dim} R=\prod_{\alpha>0}
{<\alpha,\Lambda+\delta>\over <\alpha,\delta>}, &
\end{eqnarray}
where $\delta=(\sum_{\alpha>0} \alpha)/2.$

The tensor product of irreducible representations
breaks up into a sum of irreducible representations.
In particular, for the $SU(n)$ algebras, the Young's
tableaux method allows one to compute the sum
in a straightforward manner.  For other algebras,
there are methods to perform the computations and
in particular,
the Dynkin labels also allow one to figure out
the product representations.

A final application of 
the Dynkin labels allow us to study the
branching rules of a representation under its
subalgebras.  The branching rules for
many interesting groups are catalogued in
the primary sources.


\newpage

\begin{thebibliography}{abcdefg}

\bibitem{gsw} 
S. Glashow,  Nucl. Phys. 22 (1961) 579; S. Weinberg, Phys. Rev. Lett.
19 (1967) 1264; A. Salam in {\it Elementary Particle Theory},
ed. N. Svartholm, Almqvist and Wilsell, Stockholm, 1969, p. 367.

\bibitem{cl}
For a comprehensive
discussion see, e.g., T-P. Cheng and L-F. Li,
{\it Gauge theory of elementary particle physics},
Clarendon Press, Oxford, UK, 1984.

\bibitem{azee} A. Zee, ed., {\it Unity of Forces in the
Universe,} Vol. 1, World Scientific Pub. Co. Pte. Ltd.,
Singapore, 1982.

\bibitem{gg} H. Georgi and S. L. Glashow,  Phys. Rev. Lett. 
32 (1974) 438, reprinted in Ref.\cite{azee}.

\bibitem{ps} J. C. Pati and A. Salam, Phys. Rev. D 10
(1974) 275, reprinted in Ref.\cite{azee}.

\bibitem{fm} H. Fritzsch and P. Minkowski, Ann. Phys. 93 
(1975) 193, reprinted in Ref.\cite{azee}.

\bibitem{mj} For a collection of reports, see M. Jacob, ed.,
{\it Supersymmetry and Supergravity, A Reprint Volume of
Physics Reports}, North-Holland/World Scientific, Amsterdam/Singapore,
1986.

\bibitem{hpn} H-P. Nilles, Phys. Rep. 110 (1984) 1, reprinted
in Ref.\cite{mj}.

\bibitem{ms} See, e.g., M. Sohnius, Phys. Rep. 128 (1985) 39,
reprinted in Ref.\cite{mj}. 

\bibitem{ggr} See, e.g., G. G. Ross, {\it Grand Unified 
Theories}, The Benjamin/Cummings Publishing Company, Inc.,
Menlo Park, California, USA, 1985.

\bibitem{rnm} R. N. Mohapatra, {\it Unification and
Supersymmetry}, Springer-Verlag, New York, NY, USA, 1986.

\bibitem{rnc} R. N. Cahn, {\it Semi-Simple Lie Algebras \,
and Their Representations}, 
The Benjamin/Cummings Publishing Company, Inc., 
Menlo Park, California, USA, 1984.

\bibitem{rns} R. N. Slansky, Phys. Rep. 79 (1981) 1
(reprinted in \cite{azee}).

\bibitem{hm} For a recent review, see e.g., H. Murayama, 
{\it Nucleon decay in GUT and nonGUT SUSY models},
hep-ph/9610419.  We do not discuss the proton decay
problem in this contribution and essential features
of the problem may be found in, e.g., \cite{ggr,rnm}.

\bibitem{pm} For a recent review, see e.g, P. Minkowski,
{\it  Neutrino mass and mixing}, Bern University preprint, BUTP-95/22.

\bibitem{am} B. Ananthanarayan and P. Minkowski,
{\it Status of Supersymmetric Grand Unified Theories,} preprint, hep-ph/9702279.

\bibitem{review} For some recent reviews, see, e.g.,
L. J. Hall, {\it The heavy top-quark and supersymmetry}, hep-ph/9605258;
F. Zwirner, {\it Extensions of the standard model}, hep-ph/9601300;
S. Pokorski,{\it Status of the minimal supersymmetric standard
model}, hep-ph/9510224; S. Dimopoulos, {\it Beyond the standard
model}, ICHEP 1994: 93-106 (QCD161: H51: 1994).

\bibitem{adf} U. Amaldi, W. de Boer and H. F\"urstenau,
Phys. Lett. B 260 (1991) 447; P. Langacker and M-X. Luo,
Phys. Rev. D 44 (1991) 817; 
C. Giunti, C. W. Kim and U. W. Lee, Mod. Phys. Lett. A6
(1991) 1745.

\bibitem{deBoer} For updates, see e.g., W. de Boer, 
{\it The constrained MSSM revisited}, hep-ph/9611394;
{\it Global fits to the MSSM and SM to electroweak
precision data},
hep-ph/9611395.

\bibitem{als} B. Ananthanarayan, G. Lazarides and
Q. Shafi, Phys. Rev. D 44 (1991) 1613; For a recent
update see, U. Sarid, {\it Precision top mass 
measurements vs. Yukawa unification predictions}, hep-ph/9601300.

\bibitem{sw} For a review, see, e.g., B. Schrempp and
M. Wimmer, {\it Top quark and Higgs boson masses:
interplay between infrared and ultraviolet physics}, hep-ph/9606386.

\bibitem{kubo} J. Kubo, M. Mondrag\'on and G. Zoupanos,
{\it Top quark mass predictions from gauge-Yukawa unification},
hep-ph/9512400.

\bibitem{als2} B. Ananthanarayan, G. Lazarides and Q. Shafi,
Phys. Lett.. B300 (1993) 245; B. Ananthanarayan, Q. Shafi
and X-M. Wang, Phys. Rev. D50 (1994) 5980 and references therein.

\bibitem{ap} For a comprehensive
analysis of the NMSSM with large
$\tan\beta$, see
B. Ananthanarayan and P. N. Pandita, Phys. Lett.
B 353 (1995) 70; Phys. Lett. B 371 (1996) 245.

\bibitem{abs} B. Ananthanarayan, K. S. Babu and Q. Shafi,
Nucl. Phys. B 428 (1994) 19 and references therein.

\bibitem{morereviews} For other recent directions see, e.g.,
T. Bla$\check{z}$ek et al., {\it A global $\chi^2$ analysis 
of electroweak data in SO(10) SUSY GUTs},
hep-ph/9611217; M. Carena et al., {\it Bottom-up approach and 
SUSY breaking}, hep-ph/9610341.

\bibitem{gsw2} See, e.g., M. Green, J. Schwarz and E. Witten,
{\it Superstring Theory:1, 2},
Cambridge University Press, Cambridge, 1987.

\end{thebibliography}
\end{document}